\pdfoutput=1
\documentclass[conference]{IEEEtran}
\usepackage{
    bm,
    amsmath,
    amssymb,
    booktabs,
    footnote,
    graphicx,
    hyphenat,
    multirow,
    pgfplots,
    subcaption,
    url,
    hyperref,
    siunitx,
    nicefrac,
    pifont,
    threeparttable,
    cleveref,
    adjustbox,
    enumitem,
    todonotes,
    textcomp,
    algorithm,
    algpseudocode,
}
\usepackage[utf8]{inputenc}

\author{\IEEEauthorblockN{
    Yiren Zhao\IEEEauthorrefmark{1}\textsuperscript{1}
    Xitong Gao\IEEEauthorrefmark{2}\textsuperscript{1},
    Xuan Guo\IEEEauthorrefmark{1}\textsuperscript{1},
    Junyi Liu\IEEEauthorrefmark{3},
    Erwei Wang\IEEEauthorrefmark{4}, \\
    Robert Mullins\IEEEauthorrefmark{1},
    Peter Y. K. Cheung\IEEEauthorrefmark{4},
    George Constantinides\IEEEauthorrefmark{4},
    Cheng-Zhong Xu\IEEEauthorrefmark{5}
    \IEEEauthorblockA{\IEEEauthorrefmark{1}University of Cambridge
    \{yiren.zhao, gary.guo, robert.mullins\}@cl.cam.ac.uk}
    \IEEEauthorblockA{\IEEEauthorrefmark{2} Shenzhen Institutes of Advanced Technology
    xt.gao@siat.ac.cn}
    \IEEEauthorblockA{\IEEEauthorrefmark{3}Microsoft Research Cambridge
    junyi.liu@microsoft.com}
    \IEEEauthorblockA{\IEEEauthorrefmark{4}Imperial College London
    \{erwei.wang13, p.cheung, g.constantinides\}@imperial.ac.uk}
    \IEEEauthorblockA{\IEEEauthorrefmark{5} University of Macau
    czxu@um.edu.mo}
}
}

\newcommand{\etal}{\textit{et~al\@.}}
\newcommand{\ie}{\textit{i.e\@.}}
\newcommand{\vs}{\textit{vs\@.}}

\newcommand{\tomato}{Tomato}

\newcommand{\mbnet}{MobileNet-V1}

\newcommand{\const}[1]{\mathsf{#1}}

\newcommand{\tx}{{\mathbf{x}}}

\newcommand{\ty}{{\mathbf{y}}}
\newcommand{\tz}{{\mathbf{z}}}

\newcommand{\tmu}{\bm\mu}
\newcommand{\tsigma}{\bm\sigma}
\newcommand{\tgamma}{\bm\gamma}
\newcommand{\tbeta}{\bm\beta}
\newcommand{\bnquantize}{\mathrm{quantize}_{\mathrm{8.8}}}
\newcommand{\cin}{C}
\newcommand{\cout}{C^\prime}
\newcommand{\uin}{U}
\newcommand{\uout}{U^\prime}

\newcommand\unitmu{\,\text{\textmu}}
\newcommand\unitm{\,\text{m}}
\newcommand\unitk{\,\text{k}}
\newcommand\unitM{\,\text{M}}

\newcolumntype{+}{>{\global\let\currentrowstyle\relax}}
\newcolumntype{^}{>{\currentrowstyle}}
\newcommand{\rowstyle}[1]{\gdef\currentrowstyle{#1}%
	#1\ignorespaces
}

\newcommand{\argmax}{\ensuremath\mathrm{argmax}}

\begin{document}
\title{
Automatic Generation of Multi-precision Multi-arithmetic CNN Accelerators for FPGAs}

\maketitle
\footnotetext[1]{Equal contribution.}

\begin{abstract}
    Modern deep Convolutional Neural Networks (CNNs)
    are computationally demanding,
    yet real applications often require high throughput and low latency.
    To help tackle these problems,
    we propose \tomato, a framework designed to automate the process of
    generating efficient CNN accelerators.
    The generated design is pipelined and each convolution layer
    uses different arithmetics at various precisions.
    Using \tomato, we showcase state-of-the-art multi-precision multi-arithmetic
    networks, including \mbnet{}, running on FPGAs.
    To our knowledge, this is the first multi-precision multi-arithmetic
    auto-generation framework for CNNs.
    In software,
    \tomato~fine-tunes pretrained networks to
    use a mixture of short powers-of-2 and fixed-point weights
    with a minimal loss in classification accuracy.
    The fine-tuned parameters are combined with the
    templated hardware designs to automatically produce efficient
    inference circuits in FPGAs.
    We demonstrate how our approach significantly reduces model sizes and
    computation complexities,
    and permits us to pack a complete ImageNet network
    onto a single FPGA without accessing off-chip memories
    for the first time.
    Furthermore, we show how \tomato~
    produces implementations of
    networks with various sizes running on single or multiple FPGAs.
    To the best of our knowledge, our automatically generated accelerators
    outperform closest FPGA-based competitors by at least $2$-$4\times$
    for lantency and throughput;
    the generated accelerator runs
    ImageNet classification at a rate of more than 3000 frames per second.
\end{abstract}


\section{Introduction}\label{sec:intro}

Large-scale Convolution Neural Networks (CNNs) have
delivered revolutionary performance gains to
vision applications such as
image classification~\cite{alexnet},
object detection~\cite{liu2016ssd},
and emotion recognition~\cite{mollahosseini2016going}.
To support such workloads,
both edge and cloud environments already
employ the parallelism offered by GPUs and have
more recently sought to optimize
latency, throughput and energy
with the use of FPGAs~\cite{
    brainwave, moss2018harp, Venieris2016fpga,
    qiu2016going, shen2018towards, zhao2018towards,
    bai2018cnn, shen2017maximizing, wang2019lutnet}
and ASICs~\cite{sarwar2016, wang2019csur}.

As CNN models are inherently redundant,
model compression is popular in making
CNN inference more efficient.
Methods such as
low precision quantization~\cite{zhou2017incremental, zhao2019focused}
and channel-wise structural pruning~\cite{he2017channel, gao2018dynamic}
directly shrink the compute and memory requirements.
These techniques
have become essential
for state-of-the-art CNN accelerators,
as they directly translate
to high throughput and low latency~\cite{guo2016angel}.
Unlike previous attempts~\cite{sarwar2016}
that unify all layers
in a single arithmetic at a unified precision,
we propose hybrid quantization
that allows a mixture of arithmetics and precisions
to minimize the effect of quantization
on CNNs task accuracies.
Each layer of the CNN can have different arithmetics at different precisions.
In software,
we implement hybrid quantization in \tomato~and automate the selection of
arithmetics and precisions for different layers of the CNN.
\tomato~then retrains the selected quantized model.

\begin{figure}[!h]
\includegraphics[width=\linewidth]{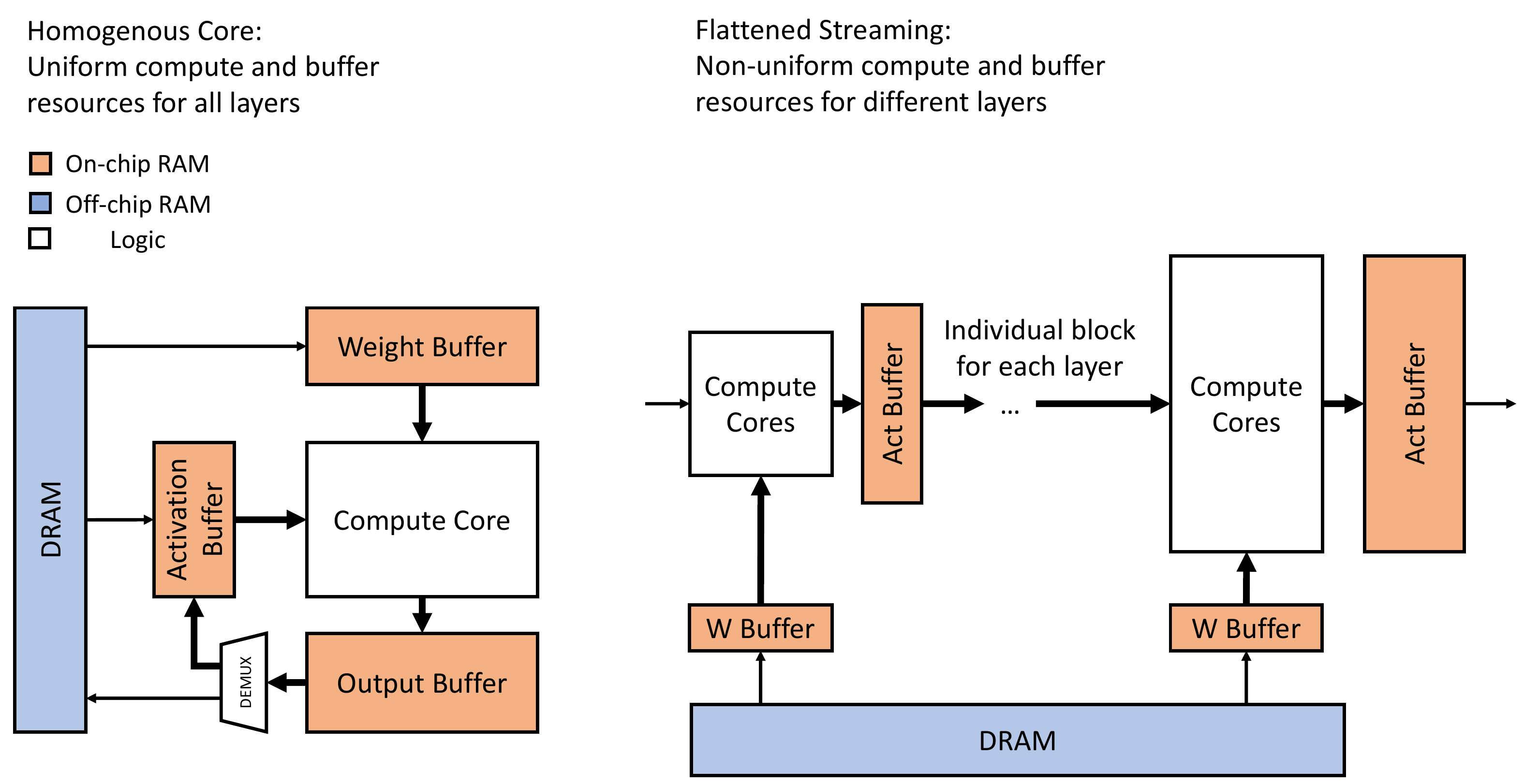}
\caption{
    An illustration of a homogenous core (left)
    and flattened streaming cores (right).
}\label{fig:arch}
\end{figure}

Hardware that uses a homogenous large systolic array
currently dominates the design of CNN accelerators;
a great number of parallel
multiplication-adds are used as a large compute core and
both weights and activations are buffered
in on-chip memory \cite{zhao2018towards, chen2016eyeriss}
(left of \Cref{fig:arch}).
The large systolic array is time-shared as a number of different
convolutions reuse the same hardware,
however, various input data sizes, channel counts, kernel sizes
and ever-emerging
new convolutions \cite{zhang2018shufflenet}
make the design of a single efficient
compute core increasingly difficult.
Alternatively, the computation of a CNN can also be divided
and pipelined into a number of smaller compute cores
(so-called Flattened Streaming).
Each computation core, streamed by activations,
is only responsible for the calculations
of individual layers \cite{brainwave, zhang2018dnnbuilder} (right of \Cref{fig:arch})
to maximize efficiency.
In this paper, we approach the CNN acceleration problem by exploiting
the reconfigurability of FPGAs in the flattened streaming design.
\tomato~directly produces small layer-wise compute cores,
maximizing the available logic of FPGAs for each target network.
We further make the observation
that flattened streaming accelerators
isolate layer-wise computations,
offering the chance
to use different arithmetics and precisions
for each layer's computation.
In addition,  the
throughputs can be matched across various compute cores
in the flattened streaming architecture --
the compute throughput of a particular layer only needs to
match its preceding layer's output generation rate.
Forcing the layers to match throughputs further reduces
the logic size of the auto-generated hardware.

The combination of hybrid quantization,
a streaming-based architecture
and ever larger FPGAs,
enable us to map all the layers
of our CNN onto a single or even multiple FPGAs\@.
In this paper, we present
an automated software-hardware co-design workflow
that produces multi-arithmetic multi-precision CNN accelerators.
The resulting hardware accelerator
is streaming-based and fully pipelined.
We make the following contributions in this paper:
\begin{itemize}[leftmargin=*]
    \item
    We demonstrate the effectiveness of hybrid quantization
    on modern efficient CNNs (like \mbnet).

    \item
    We present a novel streaming architecture for CNNs
    that is pipelined
    and uses minimal intra-layer buffering.
    Each layer's compute is matched on throughputs and
    is isolated and applied with various quatnizations.

    \item
    We show a full-stack automated workflow.
    The workflow
    packs entire networks into FPGAs.
    To the best of our knowledge,
    the resulting design outperforms
    all state-of-the-art FPGA-based CNN accelerators
    in terms of latency and throughput.
\end{itemize}

\section{Related Work}\label{sec:related}

Traditional CNNs running on GPUs typically
use single-precision floating-point arithmetic,
which is infeasible for FPGAs
with limited logic resources.
Yet CNNs are, in general,
often over-provisioned and inherently redundant;
this makes low-precision quantization
an essential technique to
drastically reduce
the memory consumed by the network's parameters,
and even allows CNN inference to be computed
entirely with low-cost arithmetic operations,
rather than floating-point ones.
Many works~\cite{
    courbariaux2014training, hubara17}
train CNNs to use low-precision weights and activations
with minimal accuracy loss,
while others pushed the limit
by using ternarized weights \( \{0, \pm1\} \)~\cite{
    hwang2014fixed, li2016ternary, zhu2016trained},
and even constraining both weights and activations
to binary values \( \{\pm1\} \)~\cite{
    Hubara2016binary, rastegari16}.
However, binarized and ternarized CNNs
struggle to achieve state-of-the-art accuracies
on large datasets.
FPGA-based accelerators generally
uniformly apply one of the above quantization methods
This specialisation provides efficiency and performance gains
when compared to GPUs with fixed set of data types.
Bit-serial accelerators \cite{sharify2019, sharma2018bit}
are also of interest as
they provide scope to optimise away superfluous computation
at the bit-level when computing with fixed-point numbers.
In contrast, the proposed hybrid quantization focuses on mixing convolution
layers with not only various precisions but also different arithmetics
in a bit-parallel manner.
Leveraging the fact that
various layers are sensitive to different quantizations,
hybrid quantization minimizes the impact of quantization loss
on the model task accuracy.

Many existing frameworks~\cite{
    guo2016angel, qiu2016going,
    shen2018towards, moss2018harp, suda2016throughput, zhang2018dnnbuilder},
that map CNN models to FPGAs
generate a large homogeneous processing core
that is temporally shared among layers.
This common design is flexible,
as by sequentially carrying out convolutions,
it is less constrained
by the amount of resources available on FPGAs.
A homogeneous core has fixed computation dimensions
which closely follows the ASIC design concept that a given
architecture is optimized for a set of chosen benchmarks \cite{shen2017maximizing}.
This approach is then challenged by the
varying size of convolutions and the emergence of new types of convolutions.
To cope with the fact that a homogeneous core is rarely optimal
for all convolution sizes and to be flexible for new convolutions,
Venieris~\etal~\cite{Venieris2016fpga}
proposed to partition a CNN model
into parts that can be separately reconfigured,
however the reconfiguration overhead
penalizes performance greatly.
Many works seek to
squeeze CNN models fully onto FPGAs,
so that they require no off-chip memory accesses
for weights and intermediate results.
Unfortunately, they are limited to
either small models~\cite{samragh17},
binarized networks~\cite{liang18},
or only a few layers of a large CNN~\cite{alwani2016fused},
which are unsuitable
for the speed and task performance on large datasets.
This paper therefore presents both the software and hardware techniques
for shrinking the resource consumption of
mapping a CNN as a flattened architecture on FPGAs.
Using the proposed framework \tomato, we demonstrate a fully pipelined \mbnet{}
--- a larger model
with over 4 million parameters ---
entirely on an FPGA\@,
which outperforms all previous designs
examined in this paper.
Furthermore, the proposed streaming-based accelerator
decouples computations in different layers. Our design, to our knowledge,
is the first multi-arithmetic and multi-precision
CNN accelerator.

\section{Hybrid Quantization}\label{sec:quantize}
Hybrid quantization mixes fixed-point quantization and shift quantization
on at a per-convolution granularity,
and all activation values between convolutions are quantized to
8-bit fixed-point numbers.

\subsection{Shift and Fixed-point Quantizations}
Using shift quantization on weights,
\ie~quantizing weights to powers-of-2 values and zeros,
avoids the costs of expensive hardware multipliers,
as they can be replaced by barrel shifters,
which results in significant savings
in terms of logic, power and latency
when compared to multipliers.
Moreover,
in the most direct hardwired implementation,
weights
simply become wiring
and can be implemented with virtually no costs.
Shift quantization results in
the following representable values,
where \( s \in \{ -1, 0, 1 \} \)
indicates the sign of the value,
\( \const{b} \) is a constant integer
shared among weights within the same layer
which ensures no values overflow,
and \( e \) is a variable exponent:
\begin{equation}
    \hat{x} = s \times 2^{e - \const{b}}.
    \label{eq:shift}
\end{equation}

The framework also allows fixed-point quantization.
An \( \const{n} \)-bit fixed-point number
with a binary point position \( \const{p} \)
can represent a value \( \hat{x} \) with:
\begin{equation}
    \hat{x} = 2^{-\const{p}} \times m_\const{n} m_{\const{n}-1} \ldots m_1,
    \label{eq:fixed}
\end{equation}

Both quantizations happen only in the feed-forward steps of the CNN.
We quantize floating-point weight values \( x \)
to the representations above,
while backpropagation bypasses the quantizations \cite{plaut1987learning}.
The pros and cons between whether to use shift or fixed-point quantizations
depend heavily on the given precision, weights distribution
and the number of values we wish to allow to saturate.
In \Cref{subsec:search}, we show how to use a greedy search
to select between different quantizations using model accuracy as the only
metric.

In addition, all ReLU activation values are constrained to
8-bit fixed-point numbers with 3-bit integers, as previously
MobileNet indicated that this does not cause a large impact on model accuracy
\cite{krishnamoorthi2018quantizing}.

\subsection{Batch Normalization}

\emph{Batch normalization} (BN) is commonly used in CNNs
to accelerate training~\cite{ioffe2015batch}.
As shown in \Cref{eq:batch_norm}, during inference,
BN normalizes convolutional outputs \( \tx \)
in a channel-wise fashion
with a moving mean \( \tmu \)
and a moving standard deviation \( \tsigma \),
then applies affine transformation on them with
the learned \( \tgamma \) and \( \tbeta \):
\begin{equation}
    \ty = \tgamma \frac{\tx - \tmu}{\tsigma} + \tbeta
    \label{eq:batch_norm}
\end{equation}
It is notable that \Cref{eq:batch_norm}
can be re-arranged into
a channel-wise affine transformation.
In the CNN feed-forward stages,
we respectively quantize
the scaling and offset factors
of this affine transformation
to fixed-point numbers:
\begin{equation}
    \ty =
        \bnquantize\left(\frac{\tgamma}{\tsigma}\right)\tx +
        \bnquantize\left(\tbeta - \frac{\tgamma \tmu}{\tsigma}\right),
    \label{eq:fused_batch_norm}
\end{equation}
where \( \bnquantize(\tz) \)
quantizes \( \tz \) into
16-bit fixed-point values
with a binary point at 8.

\subsection{Search Algorithm}\label{subsec:search}

As both the bitwidth of weights and
their representation
(i.e. either shift or fixed-width) may vary on a layer-by-layer basis,
it is intractable to explore
all possible combinations exhaustively.
For this reason,
we introduce an algorithm
which minimizes the hardware complexity
under a given accuracy constraint.
\Cref{alg:automate} provides an overview
of our search algorithm,
which accepts as inputs
a CNN model with weight parameters \( \theta \)
and \( N \) layers \( \{ l_1, l_2, \ldots, l_N \} \),
the accuracy constraint \( \alpha_\mathrm{budget} \),
the hardware resource constraint \( h_\mathrm{budget} \),
and an initial state of quantization hyperparameters \( q_0 \)
which uses 8-bit fixed-point quantization for all layers.
Here, \( \theta^\prime, \alpha \gets \mathrm{finetune} (\theta, q, E) \)
fine-tunes the model parameters \( \theta \)
under hyperparameters \( q \) for \( E \) epochs
and returns the validation accuracy of fine-tuned model.
We found empirically \( E = 3 \) is sufficient
to recover most accuracy loss due to quantization.
To traverse the search space efficiently,
we introduce a relation \( R(L) \),
where \( L \) is a set of modifiable layers.
Each transition \( (q, q^\prime) \in R(L) \)
finds a one step change
to the configuration \( q \),
\ie~decreasing the bit-width by 1
or changing the arithmetic
used by a layer \( \mathrm{layer\_changed}(q, q^\prime) \in L \).
In each step,
the algorithm is designed
to greedily find a new configuration \( q^\prime \) from \( q \)
which results in the steepest reduction in hardware resources
\( \mathrm{hwcost}(q) - \mathrm{hwcost}(q^\prime) \)
until all layers cannot be modified further
without violating the accuracy constraint.
Additionally, if the hardware resource constraint
\( \mathrm{hwcost}(q) \leq h_\mathrm{budget} \) is already satisfied
then we exit early to minimize accuracy loss.
In our experiments,
we chose \( \alpha_\mathrm{budget} \) to be \( 0.95\alpha \),
where \( \alpha \) is the original accuracy,
to generate a fully quantized model
with efficient hardware usage.
The resulting model is then fine-tuned to further increase accuracy.
\begin{algorithm}[!h]
    \caption{Search Algorithm}\label{alg:automate}
    \begin{algorithmic}[1]\small
        \Function{Search}{%
            $\theta, q_0, \alpha_\mathrm{budget}, h_\mathrm{budget}, E$}
        \State{$q \gets q_0$; $L \gets \{ l_1, l_2, \ldots, l_N \}$}
        \While{$L \neq \varnothing$}
            \State{%
                $q^\prime \gets
                    \argmax_{(q, q^\prime) \in R(L)}{
                        \left(
                            \mathrm{hwcost}(q) -
                            \mathrm{hwcost}(q^\prime)
                        \right)
                    }$
            }
            \State{$
                \theta^\prime, \alpha \gets
                    \mathrm{finetune} \left( \theta, q^\prime, E \right)
            $}
            \If{$\alpha \geq \alpha_\mathrm{budget}$}
                \State{$q \gets q^\prime$}
                \State{$\theta \gets \theta^\prime$}
                \If{$\mathrm{hwcost}(q^\prime) \leq h_\mathrm{budget}$}
                    \State{\textbf{break}}
                \EndIf{}
            \Else{}
                \State{%
                    $L \gets
                        L - \mathrm{layer\_changed}(q, q^\prime)$}
            \EndIf{}
        \EndWhile{}
        \State{\Return{$q, \theta$}}
        \EndFunction{}
    \end{algorithmic}
\end{algorithm}

\section{The Auto-generation Framework} \label{sec:hardware}
The auto-generation framework, \tomato, applies to all CNNs.
For ease of presentation, in this section,
we use the \mbnet{}
network to showcase our results
when compared to other published FPGA accelerators.

\begin{figure}[hb]
    \centering
    \includegraphics[width=0.85\linewidth]{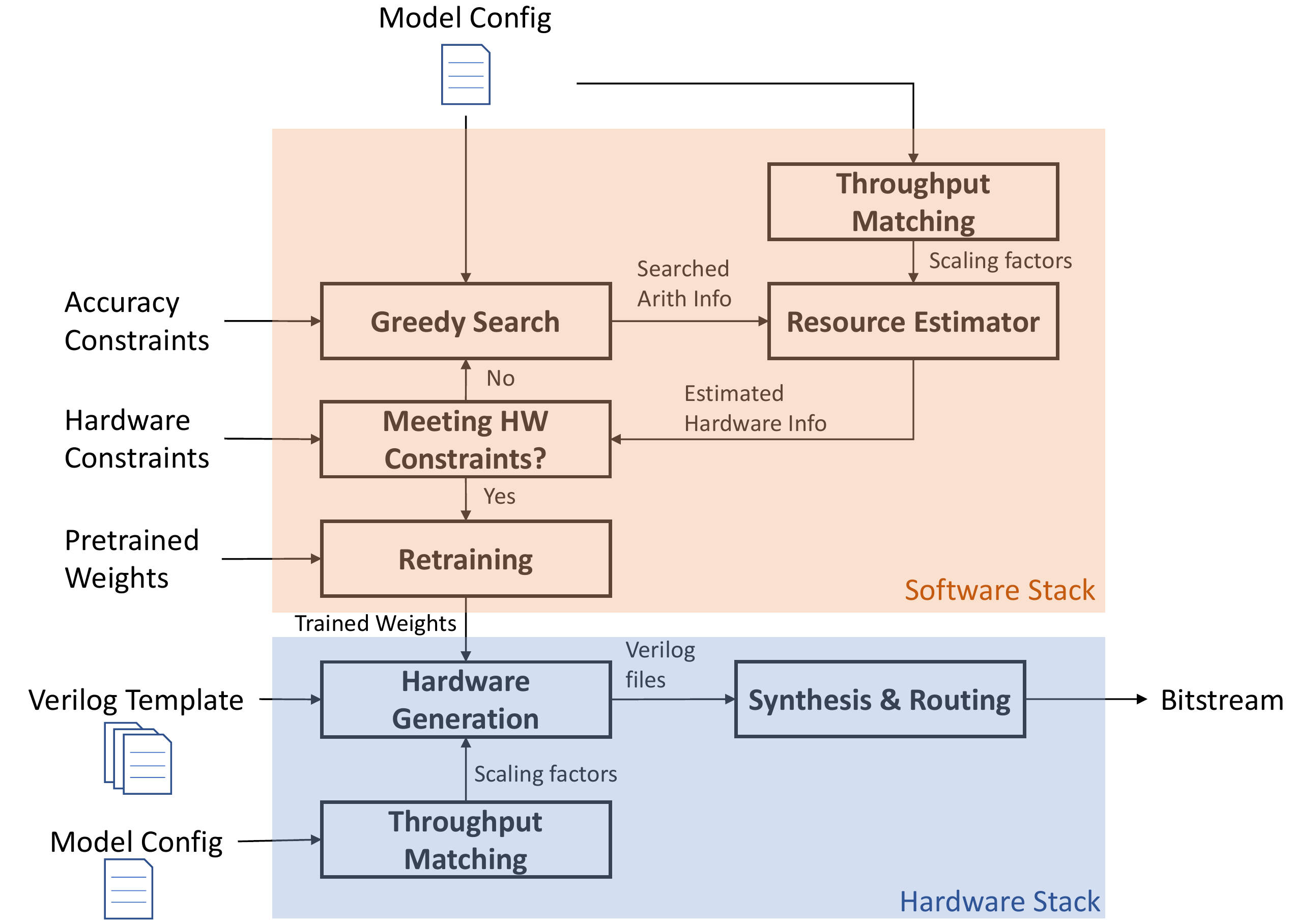}
    \caption{Framework overview for generating the accelerator for
        a targeting network on a particular dataset.
    }\label[figure]{fig:framework_overview}
\end{figure}

\subsection{Framework Overview}\label{subsec:overview}

\Cref{fig:framework_overview}
shows an overview of \tomato.
The framework starts with
an automated design process in software
which uses the algorithm in \Cref{subsec:search}
to explore the choices
of fixed-point and shift arithmetics with varying precisions
on the pre-trained CNN model.
It then produces optimized models
that are fully quantized
while satisfying the accuracy constraints.
In the exploration procedure,
it iteratively uses an accurate hardware resource estimator
to provide fast statistics of the hardware costs
and minimize the costs for the searched models.
The cost (latency, LUT and BRAM usage)
is estimated using analytical models
generated from post synthesis results
for a wide range of module parameters
The final optimized CNN model
is then fine-tuned on the original training dataset
to minimize accuracy degradation.

It is notable that from the optimized model,
\tomato~generates dedicated compute engines
for each convolutional layer.
As we have mentioned earlier,
the compute engines are connected in a pipeline,
each takes a stream of inputs
and produces a stream of outputs.
The isolated compute engines
can thus have the freedom
to use different quantizations with individual bitwidths.
To minimize hardware utilization,
layers that exceed throughput requirements
can be folded (i.e. only partially unrolled)
to share individual processing elements temporally.
In \Cref{sec:roll_unroll} we explain
how each layer can temporally share its resources,
and how we design
the throughput matching block (\Cref{sec:min_buffer})
to automatically compute
the optimal unroll factors
required to parallelize each layer
which minimizes stalls and idle circuits.
Finally, the framework generates
SystemVerilog output describing the
hardware implementation of the input model,
which is in turn synthesized into circuits
with fine-tuned weights.

\subsection{Macro-Architecture}

\begin{figure*}[!ht]
    \centering
    \includegraphics[width=0.7\linewidth]{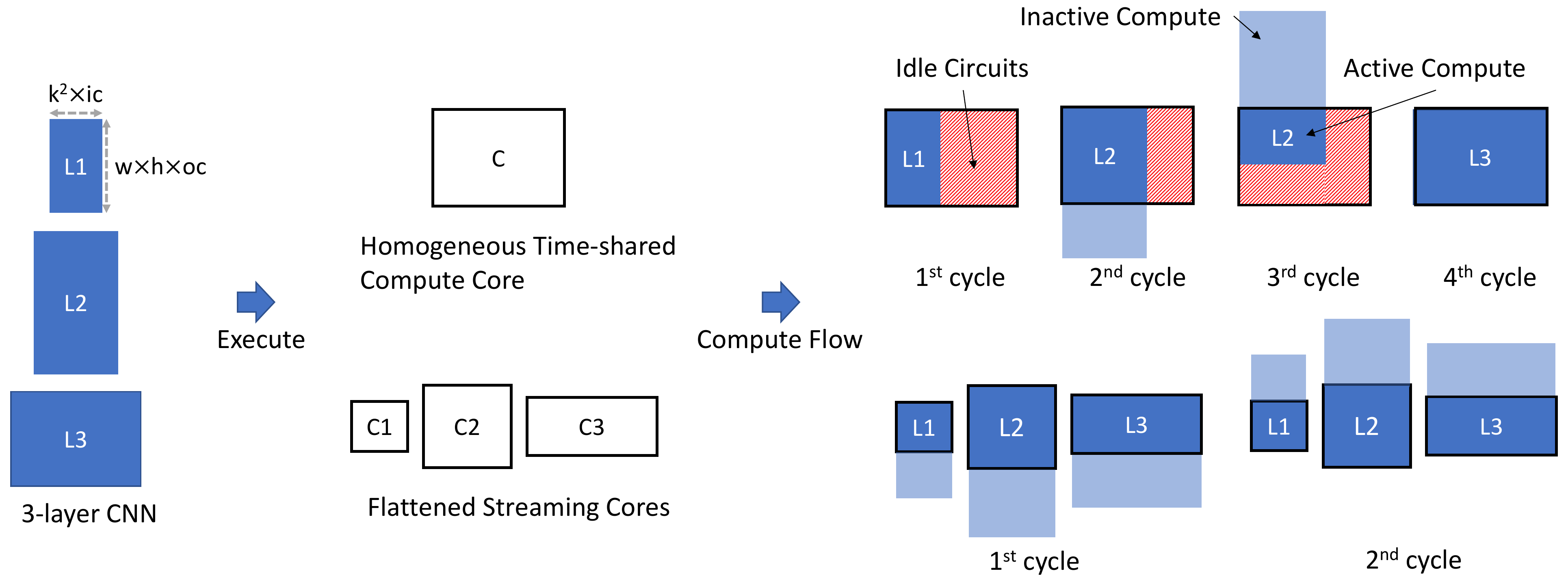}
    \caption{%
    An illustration of computation flows on executing
    three layers of convolutions ($L_1, L_2, L_3$) at different clock cycles.
    $C$ represent a large homogeneous compute core and $C_1, C_2, C_3$
    are smaller compute cores in a flattened streaming architecture.
    The rectangle block of each convolution layer represents input
    dimensions of a convolution flattened in 2D.
    \textit{k}, \textit{ic}, \textit{oc}, \textit{w}, \textit{h}
    are kernel size, input/output channels, width and height of input feature feature maps respectively.
    }\label[figure]{fig:arch_overview}
\end{figure*}
\Cref{fig:arch_overview}
shows the architecture differences
between a normal homogeneous core style accelerator
and our generated flattened streaming cores.
In the flattened streaming cores,
each convolution has its dedicated
compute engine, slide buffer and weight buffer.
Since the hardware is generated solely for the
targeted CNN and each compute core is dedicated for a particular layer,
with a suitable strategy to parallelize compute,
the generated hardware can
reach very high compute efficiency
and have minimal idle hardware.
In fact, in our measurements,
compute unit utilization is constantly high at around $84\%$.

We employ barrel shifters
or short fixed-point multipliers
in the convolution compute engines.
The weights are packed into BRAMs,
and streamed into the convolution compute engines.
Since weights are quantized
as low precision shift or fixed-point values,
the shorter bitwidths directly translates into lower BRAM usage.
Additionally, because memory ports can be time-shared,
this in turn reduces the number of BRAMs required.

\subsection{Micro-Architecture: Roll-Unrolled Convolutions}%
\label{sec:roll_unroll}

In this section,
we introduce the roll-unrolled convolution compute core,
this is designed to
minimize hardware costs
when input and output data rates permit.
As an example,
we consider a convolution layer
with a kernel size of \( K \),
which takes as input feature maps \( \tx \)
of shape \( H \times W \times \cin \),
and produces output feature maps \( \ty \)
of shape \( H^\prime \times W^\prime \times \cout \)
with \( H^\prime \) and \( W^\prime \)
depending on the stride size and padding length.
In addition, it is noteworthy
that a convolution with a stride size of 1
can produce pixels in the output feature maps
at the same rate of it taking input pixels.
A convolution with a stride size of 2, however,
produces an output pixel 4 times slower
than it can consume an input pixel.
Layers in a convolutional network
can therefore process their feature maps
at an exponentially slower rate
as more proceeding layers are strided,
and in turn have greater opportunities
to reuse data-paths.
By way of illustration,
assuming the input image
is fed at a rate of 1 cycle per pixel,
the input/output throughput rates of each layer in a \mbnet{}
can be found in the last column of \Cref{tab:unroll}.

\begin{figure}[!h]
\includegraphics[width=\linewidth]{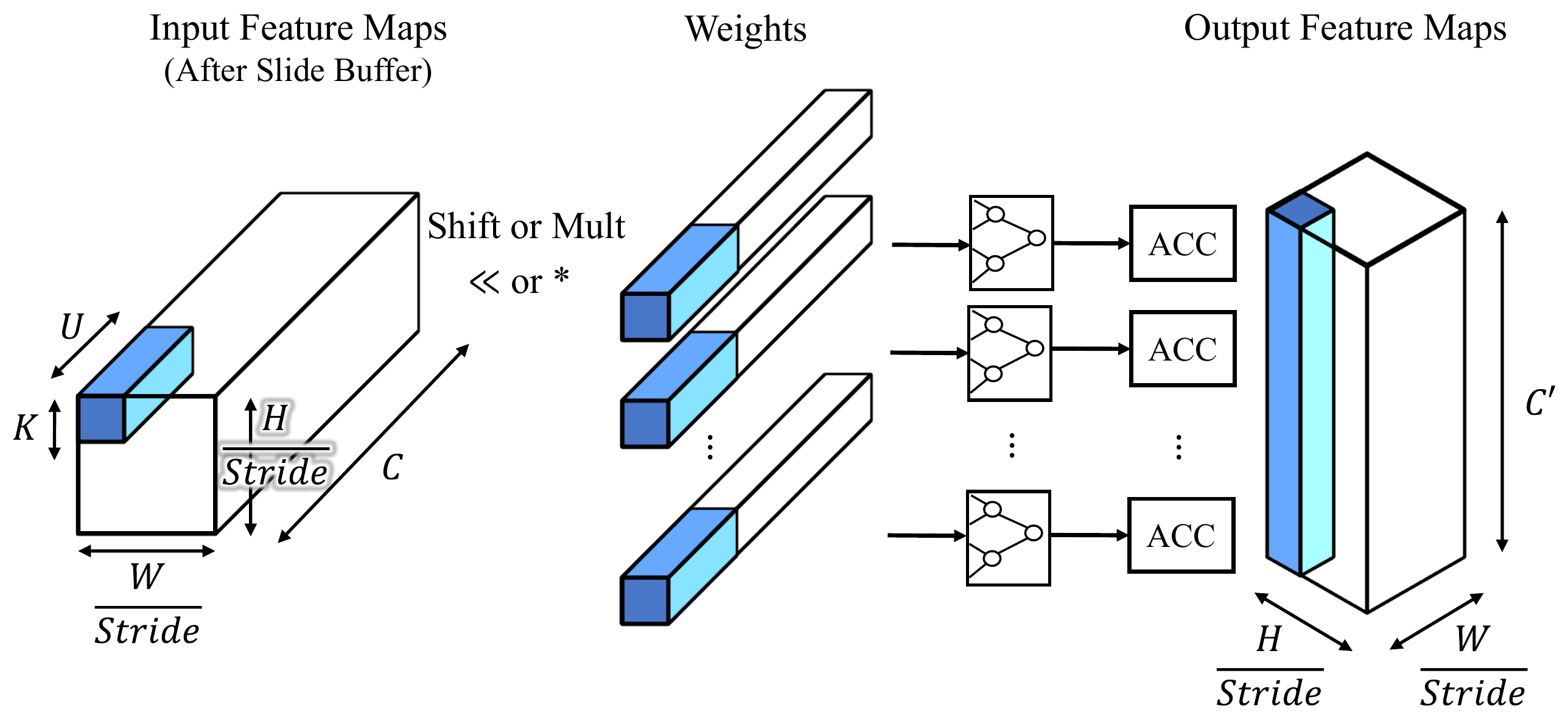}
\caption{
    An illustration of roll-unrolled computation for
    normal convolution (including pointwise convolution):
    blue indicates data elements computed in a single cycle.}
\label{fig:pointiwse}
\end{figure}

In order to maximize a layer's
utilization and minimize hardware costs,
rather than introducing stall cycles,
we introduce two unroll factors, $\uin$ and $\uout$,
for input and output channels respectively.
We partially roll
input channel dimension \( \cin \) into
$\uin$-sized blocks to
save hardware resource.
We still accumulate $\cout$ values in parallel for $\ty$.
In other words, all $\cout$ channels of
a pixel of the output feature maps
are unrolled and computed concurrently.
Fully unrolling output channels during multiplication and accumulation
is essential to allow stall-free computation.
Finally, output channels are rolled after accumulation to $\uout$-sized blocks
for batch normalization.
Fused batch normalization and ReLU operations
are time-shared for \( \uout \) output channels,
as the next layer has
an input block size equal to \( \uout \).
As we process all input channels \( \cin \)
in blocks of size \( \uin \),
we use only \( \uin \times \cout \),
instead of \( \cin \times \cout \)
parallel shift-accumulate or multiply-accumulate units,
requiring \( \left\lceil\frac{\cin}{\uin} \right\rceil \)
cycles to complete the computation
of a single pixel of all output channels,
as shown in \Cref{fig:pointiwse}.

\begin{figure}[!h]
    \includegraphics[width=\linewidth]{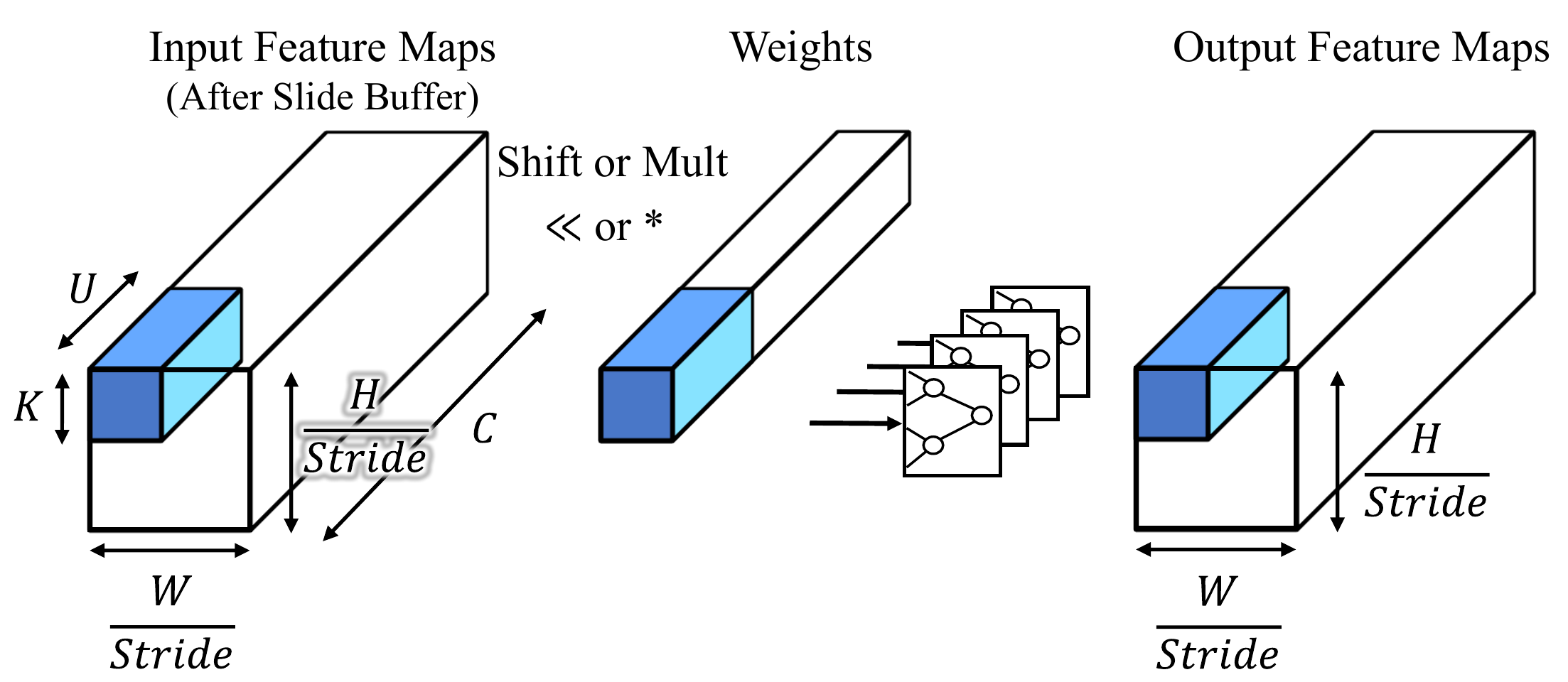}
    \caption{
    An illustration of depthwise convolution:
    blue indicates data elements computed in a single cycle.}
    \label{fig:depthwise}
\end{figure}

\tomato~does not
use roll-unrolled in depthwise convolutions.
\Cref{fig:depthwise} shows
the computation pattern for depthwise convolutions.
In contrast to normal convolutions,
depthwise convolutions are channel-wise operations,
\ie~they do not exchange information across channels.
By rolling input channels in depthwise layers,
the generated outputs are also rolled, different from
the normal roll-unrolled compute pattern.
In this way,
we exactly match the throughput
of incoming and upcoming computations
while minimizing resource utilization.
Each parallel adder tree sums up
\( K^2 \) values and is fully pipelined,
where \( K \) is the kernel size.

Roll-unrolled should not be confused with loop tiling.
Loop tiling reorders the access pattern so that it is more friendly to CPU caches and DRAM bandwidth utilization in systolic array based CNN accelerators.
As \tomato{} pipelines multiple frames instead of batching them, we did not change the access pattern.
The purpose of rolling and unrolling in \tomato{}'s streaming cores are to
minimize hardware and
provides a stall-free computation dataflow,
a fundamentally different objective.


\subsection{Striding and Rolling: Matching the Throughput}%
\label{sec:min_buffer}

By adjusting the unroll factors \( \uin \) and \( \uout \),
the framework smartly matches the throughput
between convolution layers
with different channel counts
and stridings for higher efficiency.
The only free parameter now is the input pixel rate at the
very first layer of the CNN.
The input pixel rate determines how many pixels
of an input image are fed into the accelerator at each clock cycle.
For instance, an input rate of $\frac{1}{32}$ means
we consume $1$ input pixel in $32$ clock cycles.
The choice of the input pixel rate directly impacts
the trade-off between performance and the hardware
resources required.
If this input pixel rate is $1$,
the generated hardware is optimized for performance,
fully-pipelined, and never stalls the input pixel steam.
When the input pixel rate decreases, because
of the automatic matching, unroll factors
of all subsequent convolution layers decrease and the generated
hardware thus utilizes
fewer resources but has an increased latency.

\begin{table}[ht]
\centering
\caption{
    Unrolling factors
    \( \uin \) and \( \uout \) are
    generated by the throughput matcher for \mbnet{}, depending
    on the input and output channel counts
    (\( \cin \), \( \cout \)),
    and the stride of each convolution.
    dw and pw are depthwise and pointwise convolution.
    s1 and s2 represents strides are 1 and 2.
}\label{tab:unroll}
\footnotesize
\begin{tabular}{l|lll}
\toprule
Types & \( \mathbf{\cin\text{ / }\cout} \) & \( \mathbf{\uin\text{ / }\uout} \) & \( \mathbf{\frac{\cin}{\uin}\text{ / }\frac{\cout}{\uout}} \)      \\
\midrule
 Conv / s2     & 3 / 32      & 3 / 8  & 1 / 4       \\
 Conv dw / s1  & 32 / 32     & 8 / 8  & 4 / 4       \\
 Conv pw / s1     & 32 / 64     & 8 / 16 & 4 / 4       \\
 Conv dw / s2  & 64 / 64     & 16 / 4 & 4 / 16      \\
 Conv pw / s1     & 64 / 128    & 4 / 8  & 16 / 16     \\
 Conv dw / s1  & 128 / 128   & 8 / 8  & 16 / 16     \\
 Conv pw / s1     & 128 / 128   & 8 / 8  & 16 / 16     \\
 Conv dw / s2  & 128 / 128   & 8 / 2  & 16 / 64     \\
 Conv pw / s1     & 128 / 256   & 2 / 4  & 64 / 64     \\
 Conv dw / s1  & 256 / 256   & 4 / 4  & 64 / 64     \\
 Conv pw / s1     & 256 / 256   & 4 / 4  & 64 / 64     \\
 Conv dw / s2  & 256 / 256   & 4 / 1  & 64 / 256    \\
 Conv pw / s1     & 256 / 512   & 1 / 2  & 256 / 256   \\
 Conv dw / s1  & 512 / 512   & 2 / 2  & 256 / 256   \\
 Conv pw / s1     & 512 / 512   & 2 / 2  & 256 / 256   \\
 Conv dw / s2  & 512 / 512   & 2 / 1  & 256 / 512   \\
 Conv pw / s1     & 512 / 1024  & 1 / 1  & 512 / 1024  \\
 Conv dw / s1  & 1024 / 1024 & 1 / 1  & 1024 / 1024 \\
 Conv pw / s1     & 1024 / 1024 & 1 / 1  & 1024 / 1024 \\
 Avg Pool / s1 & 1024 / 1024 & 1 / 1  & 1024 / 1024 \\
 FC / s1       & 1024 / 1000 & 1 / 1  & 1024 / 1000 \\
\bottomrule
\end{tabular}
\end{table}

We now explain how the automated throughput matching works.
The framework utilizes the classic sliding window design ---
one pixel of a output feature map
is produced once all pixels of
the sliding window on
input feature maps have arrived~\cite{bai2018cnn}.
The input stream and output stream of
strided convolutions, however, can have
different input and output rates.
For instance, when the stride size is 2,
the output stream is then 4 times
slower than the input stream
(striding occurs in the two spatial dimensions).
\Cref{tab:unroll} shows the unrolling factors
\( \uin \) and \( \uout \)
that the framework picked for each convolution in \mbnet~when choosing
the input pixel rate to be $1$.
Here, for each pixel,
\( \frac{\cin}{\uin} \)
represents the number of clock cycles
required to iterate over all input channel values,
and \( \frac{\cout}{\uout} \)
is the number of cycles
required to finish generating
all output channel values.
Taking the second
depthwise convolution layer as an example,
this layer has a stride size of 2
and the framework rolls computations
on the output channel side by a factor
of \( \frac{\cout}{\uout} = 16 \)
so that \( \uout = 4 \) values of each output
channel are computed concurrently.
Finally,
all of the unrolling information
is provided to the hardware templates
in order to instantiate the appropriate hardware.

\subsection{Batch Normalization and Rounding}

Each convolved output has an inflated precision
as mentioned in \Cref{sec:roll_unroll},
and we subsequently apply BN on these values.
As mentioned in \Cref{sec:quantize},
BN is fused and quantized
to become a channel-wise affine transformation
with fixed-point arithmetic.
We therefore use the on-chip DSP elements to perform
fixed-point multiplications and rounding after BN\@.
Since we roll computations
in output channel dimensions,
the number of multiplications
required by BN is also
significantly reduced by sharing.
It is notable that
weights share a layer-wise bias value (\Cref{eq:shift}, \Cref{eq:fixed}).
The weights bias is included
in the rounding after BN,
as it simply moves the binary point.
The final results are then
rounded to 8-bit fixed-point values
with a 5-bit fractional width.

\section{Results}\label{sec:evaluation}
\begin{table*}[!h]
\centering
\caption{
    Summary of tested networks on the \tomato. IPP stands for input pixel rate.}\label{tab:total_util}
\adjustbox{max width=\linewidth}{%
\begin{tabular}{l|c|l|l|l|l|rrrr|l|rrr}
    \toprule
    \textbf{Network}
    & \textbf{IPP}
    & \textbf{Platform}
    & \textbf{Perf Metrics} & &
    & \textbf{LUTs} & \textbf{Registers} & \textbf{BRAMs} & \textbf{DSPs} &
    & \textbf{Top-1} & \textbf{Top-5} & \textbf{Size} \\

    \midrule
    \multirow{3}{*}{MobileNet-V1}
    & \multirow{3}{*}{1}
    & \multirow{3}{*}{Intel Stratix 10}
    & \textbf{Frequency} & 156 MHz
    & \textbf{Used}
        & 926\unitk{} & 583\unitk{} & 1430 & 297
    & \textbf{Orig.} & 70.71 & 89.53 & 33.92\unitM{B} \\
    & &
    & \textbf{Latency} & 358\unitmu{s} 
    & \textbf{Total}
        & 1866\unitk{} & 3732\unitk{} & 11721 & 5760
    & \textbf{Quant.} & 68.02 & 88.02 & 16.1\unitM{B} \\
    & &
    & \textbf{Throughput} & 3109 fps
    & \textbf{Ratio} & 49\% & 15\% & 12\% & 5\%
    & \( \mathbf{\Delta} \) & -2.69 & -1.51 & 2.11\( \times \) \\

    \midrule
    \multirow{3}{*}{CifarNet}
    & \multirow{3}{*}{\( \frac{1}{32} \)}
    & \multirow{3}{*}{Intel Stratix V}
    & \textbf{Frequency} & 207\unitM{Hz}
    & \textbf{Used} & 304\unitk{} & 280\unitk{} & 771 & 84
    & \textbf{Orig.} & 91.37 & 99.68 & 4.94\unitM{B} \\
    & &
    & \textbf{Latency} & 261\unitmu{s} 
    & \textbf{Total} & 469\unitk{} & 939\unitk{} & 2.56\unitk{} & 256
    & \textbf{Quant.} & 91.06 & 99.58 & 520\unitk{B} \\
    & &
    & \textbf{Throughput} & 6317 fps
    & \textbf{Ratio} & 64\% & 29\% & 30\% & 32\%
    & \( \mathbf{\Delta} \) & -0.31 & -0.10 & 9.73\( \times \) \\

    \midrule
    \multirow{3}{*}{CifarNet}
    & \multirow{3}{*}{\( \frac{1}{288} \)}
    & \multirow{3}{*}{Intel Cyclone V}
    & \textbf{Frequency} & 116\unitM{Hz}
    & \textbf{Used} & 102\unitk{} & 84.7\unitk{} & 715 & 82
    & \textbf{Orig.} & 91.37 & 99.68 & 4.94\unitM{B} \\
    & &
    & \textbf{Latency} & 4.01\unitm{s} 
    & \textbf{Total} & 227\unitk{} & 454\unitk{} & 1.22\unitk{} & 342
    & \textbf{Quant.} & 91.06 & 99.58 & 520\unitk{B} \\
    & &
    & \textbf{Throughput} & 393 fps
    & \textbf{Ratio} & 44\% & 18\% & 58\% & 24\%
    & \( \mathbf{\Delta} \) & -0.31 & -0.10 & 9.73\( \times \) \\

    \midrule
    \multirow{3}{*}{FashionNet}
    & \multirow{3}{*}{\( \frac{1}{9} \)}
    & \multirow{3}{*}{Xilinx Artix 7}
    & \textbf{Frequency} & 98\unitM{Hz}
    & \textbf{Used} & 49.3\unitk{} & 32.7\unitk{} & 40 & 240
    & \textbf{Orig.} & 91.79 & 99.67 & 443\unitk{B} \\
    & &
    & \textbf{Latency} & 138\unitmu{s} 
    & \textbf{Total} & 63.4\unitk{} & 127\unitk{} & 135 & 240
    & \textbf{Quant.} & 91.57 & 99.56 & 65.3\unitk{B} \\
    & &
    & \textbf{Throughput} & 13.9\unitk{fps}
    & \textbf{Ratio} & 77\% & 25\% & 29\% & 100\%
    & \( \mathbf{\Delta} \) & -0.22 & -0.11 & 6.78\( \times \) \\

    \bottomrule
\end{tabular}}
\end{table*}

\subsection{Implementation Setup}

In this paper, we report results for automatically
generated hardware implementations for three distinct networks,
each optimized for a different dataset.
We use CifarNet \cite{zhao2018mayo},
an 8-layer CNN with 1.30\unitM{} parameters
and 174\unitM{} multiply-accumulates
on the CIFAR-10 dataset \cite{krizhevsky2014cifar},
\mbnet{} \cite{howard2017mobilenets}
on the ImageNet dataset \cite{imagenet}
and a customized 5-layer CNN (FashionNet)
for the Fashion MNIST dataset \cite{xiao2017fashion}.
The first two networks are relatively large,
but the last network is small.
We use \mbnet{} design as
a comparison to showcase the performance achieved from
this hardware and software co-design workflow in comparison
to other published accelerators.
The hardware part
(SystemVerilog output) is generated automatically
using templates by the Tomato framework.
We use Synplify Premier DP for synthesis and post-synthesis timing analysis.
We verified that our designs are actually implementable on FPGA by using
Xilinx Vivado to place and route the full-size MobileNet design.

\subsection{Resource Utilization}

For MobileNet,
our design is fully-pipelined
and never stalls the input stream
($\frac{\text{\#OPs}}{\text{\#OPs/cycle}}=224\times 224$).
Note that $224\times 224$ is the input image size and
this means the accelerator consumes an entire
image in exactly $224\times 224$ clock cycles.
We utilise 84\% of our 13479 compute units
(shift-and-add or multiply-and-add) on every clock cycle.
The high utilization rate of the hardware translates
to high activity ratio in the circuits
since most components are active all the time.
This fully quantized MobileNet found by \Cref{alg:automate}
uses 3-bit shift weights in its pointwise layers,
and fixed-point weights in its depthwise layers
with precisions ranging from 3 to 7.

\Cref{tab:total_util} shows the total amount of hardware utilized for the
generated accelerators for all networks on different devices.
We generate a MobileNet design with
the input pixel rate set to $1$ for best performance
(achieving 3109 FPS on an Intel Stratix 10).
The proposed workflow is a scalable one since we can adjust the input pixel
rate to control a trade-off
between performance and hardware utilization.
CifarNet results in \Cref{tab:total_util}
show how it is possible to target a small FPGA device (Cyclone V)
by adjusting this factor to $\frac{1}{288}$.
The results suggest a $3 \times$ reduction in LUT usage compared to
the design when the input pixel rate is set to $\frac{1}{32}$.
We also observe an
increase in latency, but part of the increase attributes
to the frequency differences running on various devices.
On the other hand, if provided with a small network (FashionNet),
the proposed framework generates hardware that classifies at a latency
of $0.14$ms on a very small FPGA device.
The quantized FashionNet uses 3-bit shift quantization
in the most resource-intensive third layer,
and the remaining layers use fixed-point weights with bit-widths from 5 to 7.
Although FashionNet is small, it is a good example of a
specialized network produced for resource constrained edge devices;
other examples include emotion recognition \cite{carrier2013fer}.

\begin{figure*}[!ht]
\centering
\begin{subfigure}[t]{0.32\linewidth}
    \includegraphics[width=\linewidth]{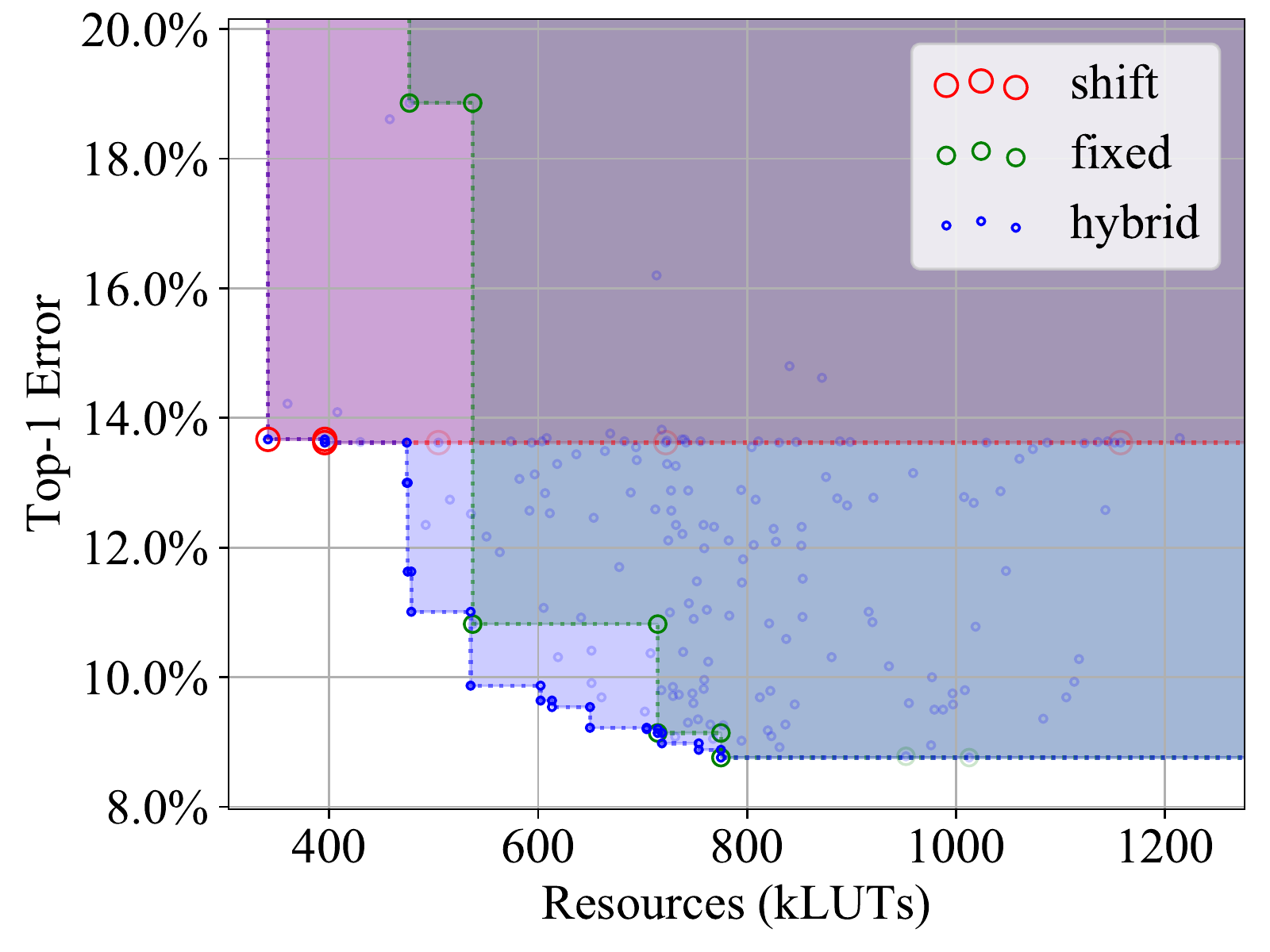}
    \caption{%
        Number of LUTs \vs{} top-1 error
        under the same \( \frac{1}{32} \) scaling
        and the same throughput.
    }\label{fig:tsf:same_latency}
\end{subfigure}~
\begin{subfigure}[t]{0.32\linewidth}
    \includegraphics[width=\linewidth]{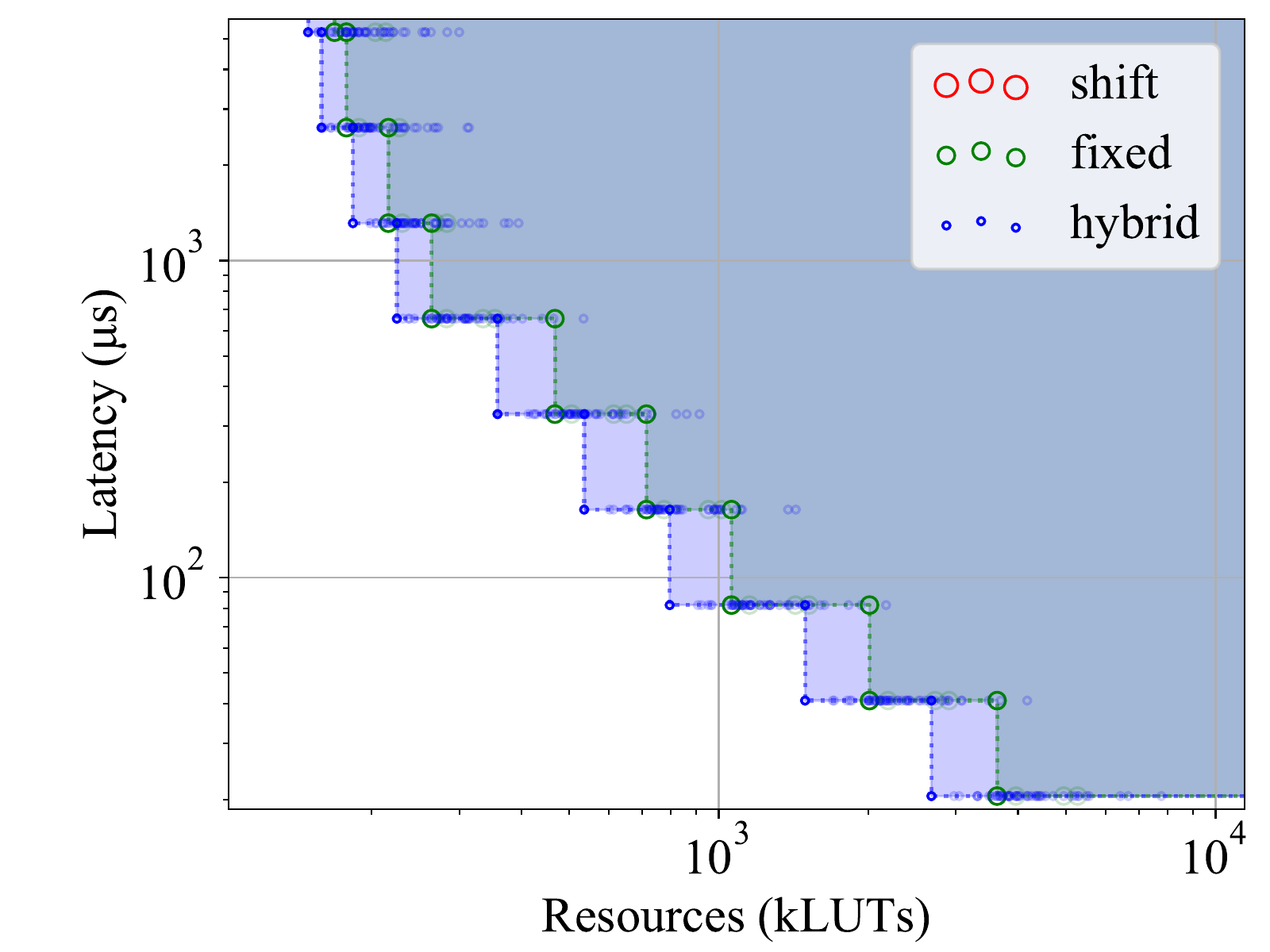}
    \caption{%
        Number of LUTs \vs{} latency
        for models with \( \leq 10\% \) top-1 errors.
    }\label{fig:tsf:lut_vs_lat}
\end{subfigure}~
\begin{subfigure}[t]{0.32\linewidth}
    \includegraphics[width=\linewidth]{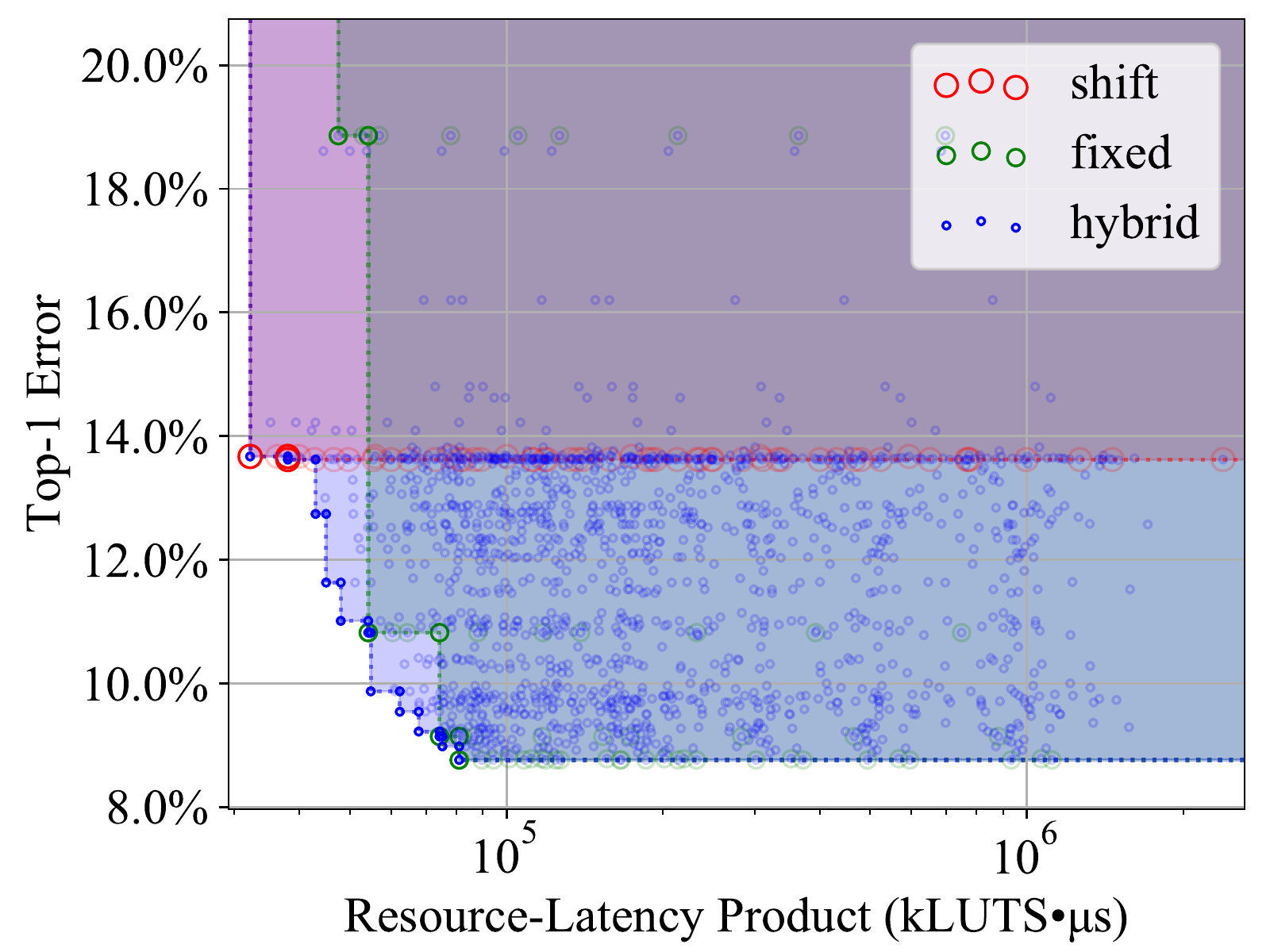}
    \caption{%
        Error \vs{} area-latency product for all optimized models.
    }\label{fig:tsf:sweep_latencies}
\end{subfigure}
\caption{%
    A case study of trade-off options
    among hardware utilization (LUTs),
    performance (latency) and
    model accuracy (top-1 error rate) before fine-tuning,
    targeting a clock frequency of \( 200 \) MHz.
    The LUTs and latency numbers are from the hardware estimator.
    Here, ``shift'' and ``fixed'' respectively
    indicate using the same
    shift and fixed-point quantization method
    across all layers with the same weight precisions.
    The ``hybrid'' points are optimized by \Cref{alg:automate}.
    The area shaded in red, green and blue
    respectively denote the 2D Pareto frontier
    of ``shift'', ``fixed'' and ``hybrid'' optimized results.
}\label{fig:tsf}
\end{figure*}

We explore in \Cref{fig:tsf}
the optimized CifarNets obtained with \Cref{alg:automate}
(denoted by the ``hybrid'' points)
and compare the results against
shift (``shift'') and fixed-point (``fixed'') models
with all layers sharing the same bit-widths ranging from 3 to 8.
To explore the trade-off
between top-1 error rates and resources,
we ran \Cref{alg:automate} \( 20 \) times
by respectively taking as inputs
the accuracy budget values \( \alpha_\mathrm{budget} \)
ranging from \( 80\% \) to \( 100\% \) at \( 1\% \) increments.
Here \( h_\mathrm{budget} \) is set to \( 0 \)
as we always minimize the resource utilization.
We constrain each layer to use either shift
or fixed-point quantized weights
and choose a bit-width ranging from 3 to 8.
Additionally, \( E = 0 \) meaning that
we skip the fine-tuning process;
without fine-tuning
the accuracies are sub-optimal
but the search process above can complete within 1 hour.
\Cref{fig:tsf:same_latency}
shows the trade-off between resource utilization and top-1 errors
under the same throughput constraints.
\Cref{fig:tsf:lut_vs_lat}
further varies the throughput scaling
of the optimized results,
and shows that when synthesized into circuits,
the the optimized models (``hybrid'')
consistently outperforms models (``shift'' and ``hybrid'')
with either shift or fixed-point quantization
under the same bit-width
applied across all layer weights.
Finally, \Cref{fig:tsf:sweep_latencies}
illustrates all results found by the three methods
and the trade-off relationship
between top-1 error rates and resource-latency products.

Hybrid quantization reduces accuracy degradations but
improves model compression rates by utilizing multi-precision
multi-arithmetic convolutions.
Importantly, we consider the ImageNet \cite{imagenet}
classification task for MobileNet.
This challenging dataset leaves less headroom for
compression techniques.
The classification results achieved on this large dataset using a
relatively compact network proves that the workflow is also robust
on other cases where the model is over-provisioned on the target dataset.


\begin{table*}[!h]
	\centering
	\caption{
		A comparison of CNN inference performance on FPGA and GPU platforms.
		The quantization of
		weights and activations are on the left.
		Target platform, frequency, latency, throughput
		and arithmetic performance are on the right.
		Metrics with $^*$ are our best-case estimations
		as they are not mentioned in the original papers.
		Note VGG16 has a similar top-5 accuracy to MobileNet-V1 when neither is
		quantized, many of the works below do not report ImageNet accuracies after quantization.}
	\footnotesize
	\begin{tabular}[c]{+c^c^c^c^c^r^r^r^r}
		\toprule
		\rowstyle{\bfseries}
		&	& \multicolumn{2}{c}{\textbf{Quantisation(s)}}
		& \multirow{2}[4]{*}{Platform}
		& {\multirow{2}[4]{*}{\shortstack{Frequency\\(MHz)}}}
		&
		{\multirow{2}[4]{*}{\shortstack{Latency\\(ms)}}}
		& {\multirow{2}[4]{*}{\shortstack{Throughput\\(FPS)}}}
		& {\multirow{2}[4]{*}{\shortstack{Arithmetic\\perf. (GOP/s)}}} \\
		\cmidrule{3-4}
		\rowstyle{\bfseries}
		&Implementation & Weights	& Acts	\\
		\midrule
		\multirow{7}{*}{\rotatebox{90}{\shortstack{VGG16}}}
		& Throughput-Opt~\cite{suda2016throughput}	& FXP8	& FXP16	& Intel Stratix V & 120	& 262.9 & 3.8$^*$	& 117.8	\\
		& fpgaConvNet~\cite{Venieris2016fpga}	& FXP16	& FXP16	& Xilinx Zynq XC7Z045 & 125	& 197$^*$ & 5.07	& 156	\\
		& Angel-Eye~\cite{guo2016angel}	& BFP8	& BFP8	& Xilinx Zynq XC7Z045 & 150 & 163$^*$ & 6.12$^*$	& 188	\\
		& Going Deeper~\cite{qiu2016going}	& FXP16	& FXP16	& Xilinx Zynq XC7Z045	& 150 & 224$^*$ & 4.45	& 137	\\
		& Shen \emph{et al.}~\cite{shen2018towards}	& FXP16	& FXP16	& Xilinx Virtex US XCVU440 & 200  & 49.1	& 26.7	& 821	\\
		& HARPv2~\cite{moss2018harp}	& BIN	& BIN	& Intel HARPv2	& {--} & 8.77$^*$	& 114	& 3500	\\
		& GPU~\cite{moss2018harp}	& FP32	& FP32	& Nvidia Titan X	& {--} & {--}	& 121	& 3590	\\
		\midrule
		\multirow{5}{*}{\rotatebox{90}{\shortstack{MobileNet}}}
		& Ours		& Mixed	& FXP8	& Intel Stratix 10	& 156	 & 0.32	& 3109	& 3536 \\
		& Ours		& Mixed	& FXP8	& Xilinx Virtex US+ XCVU9P  &125 & 0.40	& 2491	& 2833\\
		\cmidrule{2-9}
		& Zhao \emph{et al.}~\cite{zhao2018towards} & FXP16	& FXP16	& Intel Stratix V	& 200	& 0.88 & 1131	& 1287  \\
		& Zhao \emph{et al.}~\cite{Zhao2018fpga} & FXP8	& FXP8	& Intel Stratix V & 150 & 4.33 & 231	& 264	\\
		& GPU & FP32 & FP32	& Nvidia GTX 1080Ti	& {--} & 279.4 & 515 & 586\\
		\bottomrule
	\end{tabular}
	\label{tab:perf_comp}
\end{table*}

\subsection{Performance Evaluation}
\label{subsec:eval}
We compare the MobileNet-V1 design generated by our framework with
existing FPGA accelerators in \Cref{tab:perf_comp}.
This comparison only considers networks in the ImageNet dataset
that achieves greater than $70\%$ top-1 accuracy when not quantized.
The computer vision community spends a significant amount of effort in
optimizing model architecture,
we note that it is important to generate results for the latest models
as they offer the best accuracy/cost trade offs.
Results for older models in terms of GOP/s can be misleading.

Our design is different from most existing designs
examined in \Cref{tab:perf_comp}
in the following ways.
First, our framework exploits hybrid quantization to
minimize the impact of quantization errors.
Second, using the throughput scaling trick,
the amount of hardware required
is reduced significantly.
Most of the examined designs rely on a high DRAM bandwidth with
a large monolithic compute core.
As discussed previously, a large compute core cannot
explore multi-precision and multi-arithmetic quantizations
and struggles to fully utilize
compute units on convolutions with varying channel counts, kernel sizes and
input feature sizes.
\tomato~generated designs compute various layers concurrently
and quantize each layer differently,
thus achieving a very high utilization of our compute units
and to operate at around
3.5 TOPs/s on Stratix 10.
Note that,
for accelerators that we compare against,
the arithmetic performance reported
in \Cref{tab:perf_comp} considers the peak performance
assuming unbounded DRAM bandwidth.
In reality, such designs can easily be
limited by the available memory bandwidth.
In contrast,
this is not a concern in our design
as all weights and activations are held on the FPGA.
Additionally,
our design has a high throughput since operations rarely stall.
Designs proposed by
Bai \etal~\cite{bai2018cnn} and Zhao \etal~
\cite{zhao2018towards} have to
execute computations in a layer-wise fashion, and thus operations
in the next layer only executes when the current layer finishes.
In our framework, similar to Shen \etal~\cite{shen2017maximizing},
computations in different layers happen concurrently
in the same pipeline stage while later layers never stall
earlier layers.
Moreover, consecutive image inputs can be fully pipelined, because
we utilize streaming sliding windows.
These features help us to achieve high throughput compared
to other designs (\Cref{tab:perf_comp}).
The proposed workflow avoids complex and time-consuming design space
exploration
as necessary in
many compared FPGA accelerators \cite{bai2018cnn, Zhao2018fpga}.

In terms of performance, our design achieves a higher throughput
and a lower latency compared to all designs (as listed in \Cref{tab:perf_comp}).
We notice that most CNN accelerators report \textit{theoretical upper bounds}
for arithmetic performance and throughput.
In terms of latency,
the numbers are reported optimistically
assuming DRAM accesses cause no stalls.
In our design, since we stream in pixels of the input image,
the computation pattern differs from most existing designs.
The reported values in \Cref{tab:perf_comp} represents our true performance
and make no assumptions regarding DRAM bandwidth.
Our system automatically produces an implementation of MobileNet for
the Stratix 10 FPGA that outperforms
Zhao~\etal~\cite{zhao2018towards}
by \( 2.44\times \) in latency and \( 3.52\times \) in throughput.

\subsection{Multi-FPGA Acceleration}
\label{sec:multifpga}

The flattened streaming style employed by \tomato{} makes it easy to
partition the generated design across multiple FPGAs
This feature makes \tomato{} highly scalable with respect to network sizes and/or FPGA sizes.
We demonstrate in this section an example of partitioning MobileNet-V1 onto two Stratix V FPGAs, connected through enhanced small form-factor pluggable (SFP+) interfaces.
We present the performance results in comparison to Zhang \etal~\cite{zhang2016}
in \Cref{tab:multi_comp}, and the detailed hardware utilization information
in \Cref{tab:multi_util}.
The latency is not penalised thanks to the low latency of SFP+,
which contributes only a $0.0013$ms latency overhead.

The simple case study of partitioning the same MobileNet-V1 design to two
devices demonstrates that, first, \tomato~generated designs are
scalable from single to multi-FPGAs; second, aiming accelerating
new network architectures with mixed quantizations
bring significant improvements in accuracies,
latency and throughput.

\begin{table}[!ht]
\centering
\caption{
    Multi-FPGA acceleration of CNNs.
    MBNet represents MobileNet, VGG-D and VGG-E are both
    VGG16 based networks but different configurations,
    one is latency oriented and one is throughput oriented \cite{zhang2016}.
}\label{tab:multi_comp}
\vspace{-5pt}
\footnotesize
\begin{tabular}{llrrr}
\toprule
\textbf{Network} & \textbf{Acc (\%)} & \textbf{\#Device} & \textbf{Lat (ms)} & \textbf{Tpt (GOPs)} \\
\midrule
MbNet-V1 (ours)    & 68.02   & 2 Stratix V    & 0.32     & 3536     \\
VGG-D \cite{zhang2016} & 66.52   & 2 VX690t       & 200.9    & 203     \\
VGG-E \cite{zhang2016} & 66.51 & 7 VX690t       & 151.8    & 1280  \\
\bottomrule
\end{tabular}
\vspace{-5pt}
\end{table}
\begin{table}[!ht]
\centering
\caption{
    Multi-FPGA hardware utilization.
}\label{tab:multi_util}
\vspace{-5pt}
\footnotesize
\begin{tabular}{rrrrrr}
\toprule
\textbf{Device No} & \textbf{Frequency} & \textbf{LUTs} & \textbf{Regs} & \textbf{BRAM} & \textbf{DSP} \\
\midrule
0    & 156MHz   & 362.7k      & 278.8k       & 828     & 256    \\
1    & 156MHz   & 345.9k      & 303.6k       & 598     & 31     \\
\bottomrule
\end{tabular}
\vspace{-5pt}
\end{table}

\section{Conclusion}\label{sec:conclusion}
In this paper,
we presented a hardware-software
co-design workflow to automatically generate high-performance CNN accelerators.
The workflow is able to quantize weights to both fixed-point and shift values
at various precisions, and keeps activations
to fixed-point numbers.
In addition, it transforms batch normalization to
simple affine operations with fixed-point scaling and offset factors.
In hardware, the framework utilizes the Roll-Unrolled compute pattern
and provides flexibility in rolling computations in the channel dimension.
As a result, the guided rolling minimizes computation while
keeping the input stream stall-free.
The results showed state-of-the-art performance
in terms of model accuracy, latency and throughput.
The implemented accelerator for MobileNet is fully pipelined with
sub-millisecond latency (0.32ms) and is able to classify
at around 3000 frames per second.

\section*{Acknowledgments}

We thank EPSRC for providing
Yiren Zhao his doctoral scholarship.
Xitong Gao is supported
by the National Natural Science Foundation of China
(No. 61806192).

\bibliographystyle{abbrv}
\bibliography{references}

\end{document}